\documentclass[conference]{IEEEtran}
\IEEEoverridecommandlockouts
\usepackage{cite}
\usepackage{amsmath,amssymb,amsfonts}
\usepackage{algorithmic}
\usepackage{graphicx}
\usepackage{textcomp}
\usepackage{url}
\usepackage{xcolor}
\usepackage{footmisc}
\usepackage{multirow}
\usepackage{balance}
\def\BibTeX{{\rm B\kern-.05em{\sc i\kern-.025em b}\kern-.08em
    T\kern-.1667em\lower.7ex\hbox{E}\kern-.125emX}}

\definecolor{LightGray}{rgb}{0.78, 0.79, 0.85}
\definecolor{LightCyan}{rgb}{0.,0.5,0.71}

\begin{document}
\title{A Taxonomy of Inefficiencies in LLM-Generated Python
Code}




\author{
\IEEEauthorblockN{
Altaf Allah Abbassi\textsuperscript{\dag}, 
Leuson Da Silva\textsuperscript{\dag}, 
Amin Nikanjam\textsuperscript{\ddag}\thanks{\textsuperscript{\ddag}Work done while at Polytechnique Montréal.}, 
Foutse Khomh\textsuperscript{\dag}
}
\IEEEauthorblockA{
\{altaf-allah.abbassi, leuson-mario-pedro.da-silva, foutse.khomh\}@polymtl.ca, amin.nikanjam@h-partners.com
}
\IEEEauthorblockA{
\textsuperscript{\dag}Polytechnique Montréal, Montreal, Quebec, Canada\\
\textsuperscript{\ddag}Huawei Distributed Scheduling and Data Engine Lab, Canada
}
}

\maketitle

\begin{abstract}
Large Language Models (LLMs) are widely adopted for automated code generation with promising results. Although prior research has assessed LLM-generated code and identified various quality issues- such as redundancy, poor maintainability, and sub-optimal performance- a systematic understanding and categorization of these inefficiencies remain unexplored. 
Therefore, we empirically investigate inefficiencies in LLM-generated Python code by state-of-the-art models, i.e., CodeLlama, DeepSeek-Coder, and CodeGemma. 
To do so, we manually analyze 492 generated Python code snippets in the HumanEval+ dataset. 
We then construct a taxonomy of inefficiencies in LLM-generated Python code that includes 5 categories (\textit{General Logic}, \textit{Performance}, \textit{Readability}, \textit{Maintainability}, and \textit{Errors}) and 19 subcategories of inefficiencies. 
We validate the obtained taxonomy through an online survey with 58 LLM practitioners and researchers. 
The surveyed participants affirmed the completeness of the proposed taxonomy, and the relevance and the popularity of the identified code inefficiency patterns.
Our qualitative findings indicate that inefficiencies are diverse and interconnected, affecting multiple aspects of code quality, with \textit{logic} and \textit{performance-related} inefficiencies being the most frequent and often co-occurring while impacting overall code quality. 
Our taxonomy provides a structured basis for evaluating the quality of LLM-generated code and guiding future research to improve code generation efficiency.

\end{abstract}


\section{Introduction}
\label{sec: introduction}
Large Language Models (LLMs) have revolutionized software development by enabling automated code generation~\cite{lyu2024automatic, hou2024large, zheng2025towards}. In recent years, we have witnessed the emergence of LLMs specifically tailored for code generation, including open-source models such as CodeLlama~\cite{roziere2023code} and proprietary models like GPT-4~\cite{achiam2023gpt}.
These models have demonstrated the ability to generate code across multiple programming languages, including Python, Java, and C/C++~\cite{liu2024your}, showcasing their potential to assist developers in various tasks~\cite{sharma2024llms, wang2023review}.

While LLMs show promising results in code generation, their generated code, similar to human-written code, exhibit various quality issues~\cite{michelutti2024systematic, wang2023review}, which can impede their smooth 
integration into production environments. 
For instance, LLM-generated code has shown to suffer from redundancies, unnecessary computations, and suboptimal implementations~\cite{MORADIDAKHEL2023111734}, resulting in increased execution time, higher memory consumption, maintainability challenges, and import errors~\cite{siddiq2024quality, zhong2024can}. Previous studies have also shown that LLMs may generate incorrect logic, misinterpret task requirements, and struggle to handle corner cases, leading to reliability issues~\cite{tambon2024bugs}. 
Even when functionally correct, LLM-generated code can face quality issues~\cite{dou2024s}. It may be difficult to maintain~\cite{siddiq2024quality}, unnecessarily complex~\cite{MORADIDAKHEL2023111734}, prone to security vulnerabilities, and require manual intervention before deployment.  
These issues pose significant risks to software quality, particularly for developers adopting LLMs in real-world settings~\cite{dunne2024weaknesses}.

While prior research has examined specific quality aspects, such as correctness, security, and maintainability in LLM-generated code~\cite{zheng2024beyond, licorish2025comparing, dou2024s}, inefficiencies as a broader concept remain largely unexplored. 
Existing studies focus only on buggy-generated code~\cite{tambon2024bugs, dou2024s} to categorize bugs and identify root causes. 
However, this narrow focus overlooks a critical issue: \textit{even functionally correct code can exhibit inefficiencies that degrade software quality, such as performance bottlenecks~\cite{MORADIDAKHEL2023111734}, limiting real-world adoption of LLM-generated code.}

In this study, we aim to bridge this gap by identifying, categorizing, and analyzing overall inefficiency patterns, regardless of correctness, in LLM-generated code from three leading open-source LLMs for code generation. In this work, we define inefficiencies as any issues in LLM-generated code that degrade its quality, hinder integration into larger systems, or even affect functionality in isolation.
Understanding inefficiencies guides the development of more efficient LLMs and helps practitioners and researchers improve LLM-generated code. 
Our investigation is guided by the following Research Questions (RQs):
\begin{itemize}
    \item \textit{\textbf{RQ1} What inefficiency patterns occur in LLM-generated code?}
    \item \textit{\textbf{RQ2} How relevant are these inefficiencies to software practitioners and researchers?}
\end{itemize}

We conduct our study using \textit{HumanEval+}\, a widely used benchmark for code generation composed of 164 human-crafted Python tasks~\cite{liu2024your}. 
We selected three leading open-source LLM models -CodeLlama, DeepSeek-Coder, and CodeGemma- to generate code for the different tasks, resulting in 492 generated code snippets. To evaluate the quality and requirements conformity of the LLM-generated Python code, we employed the higher-capacity GPT-4o-mini model as a judge. The judgments provided initial guidelines for our manual exploratory and qualitative analysis to identify inefficiency patterns in LLM-generated code. We adopted open coding~\cite{seaman1999qualitative} to iteratively derive categories of inefficiencies. 
Our study results in a taxonomy spanning five categories: \textit{General Logic}, \textit{Performance}, \textit{Readability}, \textit{Maintainability}, and \textit{Errors}, comprising 19 subcategories. We validate this taxonomy through a survey with 58 software practitioners and researchers who use LLMs for coding assistance.

Our findings indicate that 33.54\% of the studied sample exhibited multiple inefficiencies, indicating that inefficiencies in LLM-generated code are diverse and interconnected.
\textit{General Logic} and \textit{Performance} inefficiencies are the most frequent and often co-occurring with \textit{Maintainability}, and \textit{Readability} inefficiencies. 
This co-occurrence suggests that limitations in LLMs' ability to generate correct and efficient code contribute to broader inefficiencies affecting code quality.
The surveyed participants largely validated the relevance and prevalence of identified inefficiencies, confirming that it is aligned with real-world inefficiencies in LLM-generated code regardless of the model used. 
Additionally, while participants acknowledged these inefficiencies, they emphasized that in practice, they prioritize correctness over other code aspects.
Our findings highlight a critical gap in LLMs' capability to generate correct, optimized, and high-quality code.


For LLM developers, our taxonomy offers insights into existing inefficiencies of LLM-generated code. This knowledge can inform future targeted LLMs code generation improvement directions.
For researchers, our categorization provides a structured foundation for further studies on inefficiencies in LLM-generated code. Additionally, our findings can support the development of automated detection techniques and refactoring tools targeting these inefficiencies to enhance code quality as LLM-generated code becomes increasingly adopted in production systems. 
To summarize, this paper makes the following contributions:
\begin{itemize}
    \item We study inefficiencies observed in LLM-generated Python code and their prevalence. 
    \item We propose a taxonomy of inefficiencies in LLM-generated Python code based on their characteristics. 
    \item We validate our findings using an online survey and provide insights for researchers and practitioners.
    \item We make our data and results publicly available in our replication package~\cite{ReplicationPackage2025}.

\end{itemize}

The rest of the paper is organized as follows: In Section \ref{sec: backgound}, we present the background. Next, we detail the methodology used to construct and validate the taxonomy of inefficiencies in Section \ref{sec: methodology}. In Section \ref{sec: results}, we report our empirical findings, presenting our taxonomy. Then, we discuss related work and threats to the validity of our results in Section \ref{sec: related_work} and \ref{sec: threats_validity}, respectively. Finally, Section \ref{sec: conclusion} concludes the paper. 

\section{Background}
\label{sec: backgound}
In this section, we introduce the three LLMs used for code generation in our study. 
\subsubsection*{CodeLlama} is an open-source family of models based on Llama 2~\cite{touvron2023llama} developed by Meta for code-related tasks such as completion and infilling.
It includes the foundation model (CodeLlama), a Python-specialized model (CodeLlama-Python), and an instruction-following model (CodeLlama-Instruct), available in different sizes (7B, 13B, 34B, and 70B parameters). 
CodeLlama is trained on a near-deduplicated dataset of publicly available code, it incorporates a small fraction of natural language data related to programming to enhance comprehension ~\cite{roziere2023code}.


\subsubsection*{DeepSeek-Coder} provides a range of open-source models trained from scratch on 2 trillion tokens across 87 programming languages. 
The pretraining data is organized at the repository level to enhance model capability~\cite{guo2024deepseek}.
The models employ two key techniques: fill-in-the-middle and next-token prediction. 
Just like CodeLlama, DeepSeek-Coder is also available in different sizes, ranging from 1.3B to 33B parameters, with both foundational and instruction-tuned versions. 
These models leverage a 16K context window to improve code generation and infilling~\cite{guo2024deepseek}. 

\subsubsection*{CodeGemma} is a collection of specialized open-code models based on Google DeepMind's Gemma models~\cite{team2024gemma}, trained on 500 to 1000 billion tokens, primarily for code-related tasks. 
The CodeGemma family includes: a 7B pre-trained model, a 7B instruction-tuned model, and a specialized 2B model designed for code infilling and open-ended generation. 
The 2B model is trained entirely on code, while the 7B models are trained on a mix of 80\% code and 20\% natural language, sourced from deduplicated publicly available code repositories. 
These models are trained using a fill-in-the-middle task to enhance their coding capabilities~\cite{team2024codegemma}. 

\section{Methodology}
\label{sec: methodology}
\begin{figure*}[h!]
  \centering
  \includegraphics[width=0.7\linewidth]{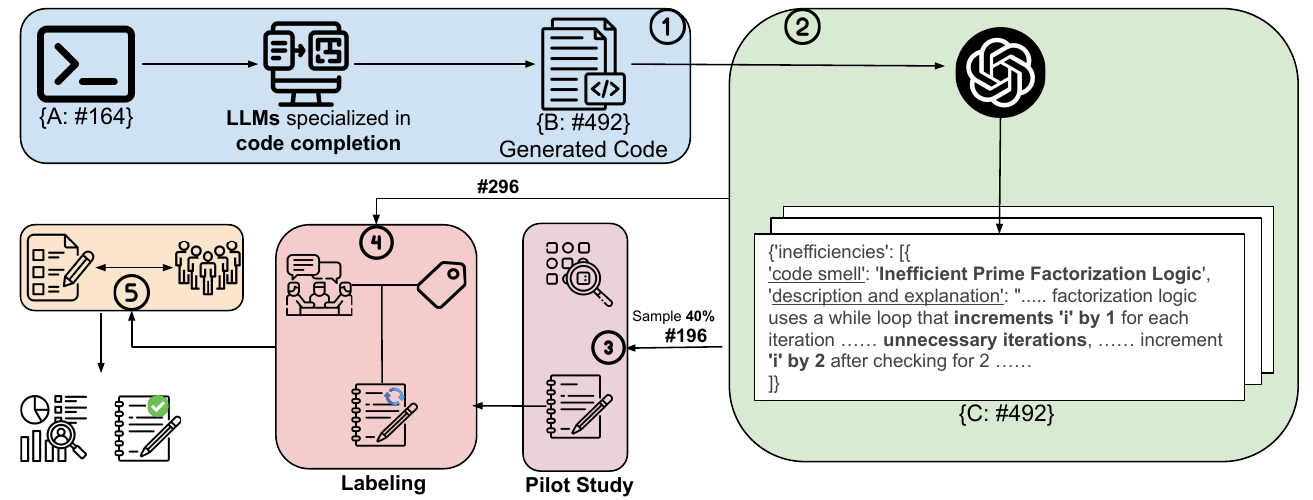}
  \caption{The methodology we followed for our study.}
  \label{fig:taxonomy_methodology}
  \vspace{-1em}
\end{figure*}

This section presents the methodology used in this study to construct and validate the taxonomy of inefficiencies in LLM-generated Python code. The process consists of five steps, as illustrated in Figure \ref{fig:taxonomy_methodology}.
First, we generated code using the HumanEval+ benchmark~\cite{liu2024your} and collected data from three open-source LLM4Code models: CodeLlama~\cite{roziere2023code}, DeepSeek-Coder~\cite{guo2024deepseek}, and CodeGemma~\cite{team2024codegemma}. Second, we leveraged a higher-capacity model (GPT-4o-mini) to judge the generated code and identify existing inefficiencies. In the third and fourth steps, we manually analyzed the code samples and the model's judgments to develop a comprehensive taxonomy. Finally, in the fifth step, we conducted a survey with practitioners and researchers to validate the LLM-generated code inefficiencies identified in our taxonomy. The details of each step are presented in the remainder of this section.

\subsection{Step 1: Data Selection and Collection}
Regarding the data collection, considering the popularity and wide adoption of Python as a programming language~\cite{spectrum2024}, this study focused on Python-generated LLM code. 
Given the focus on Python, we selected HumanEval+~\cite{liu2024your} (denoted as {$\mathit{A}$ in Figure \ref{fig:taxonomy_methodology}), a widely recognized and extensively employed benchmark for code generation~\cite{ouyang2023llm, lin2024llm, jiang2024survey, miah2024user}. HumanEval+ enhances the original HumanEval~\cite{chen2021evaluating} dataset by augmenting it with additional test cases, enabling a more rigorous and reliable assessment of the functional correctness of the generated code~\cite{liu2024your}.  
HumanEval+ comprises 146 Python programming tasks, each accompanied by (i) a task description provided as a doc string and (ii) associated unit tests to rigorously evaluate the correctness of the code snippets. 
Regarding the LLMs, we aimed to select those trained or fine-tuned for code generation tasks to enable a more meaningful and targeted analysis of inefficiencies.
To ensure accessibility and feasibility, we established two selection criteria: (i) the models must be open source and (ii) have fewer than 8B parameters.
Based on these criteria, we selected the three highest-performing models according to the HumanEval+ leaderboard~\cite{evalplusLeaderboard}: CodeLlama-7B~\cite{codellama2024}, 
CodeGemma-7B~\cite{codegemma2024}, and DeepSeek Coder-6.7B~\cite{deepseekcoder2024}.  
These models, beyond ranking among the top-performing models~\cite{evalplusLeaderboard}, are also widely adopted for code generation tasks~\cite{jiang2024survey}.
Although our selection focuses on small open-source models, these models demonstrate performance comparable to proprietary ones, surpassing GPT-3.5~\cite{guo2024deepseek} and approaching GPT-4~\cite{zheng2024opencodeinterpreter}, ensuring representativeness in our study.
After defining the setup for our study, we proceeded with generating the code snippets.
For the hyperparameters, we adopted a conservative approach by setting the temperature to $0$, ensuring deterministic outputs and minimizing randomness in code generation.
For prompt construction, we followed the default configuration recommended in the official Hugging Face documentation~\cite{codegemma2024, codellama2024, deepseekcoder2024}.
This approach ensured a fair assessment of inherent inefficiencies in LLM-generated code rather than artifacts caused by inadequate prompts.  
As a result of this step, we have a dataset of \textit{$164$} generated code snippets for each LLM, resulting in \textit{$492$} code snippets ($\mathit{B}$), from the initial dataset of tasks ($\mathit{A}$). 

\subsection{Step 2: Judging LLM-generated Code}
LLMs have been effectively used to assess code or tests~\cite{zhao2024codejudge, sollenberger2024llm4vv}, proposing a paradigm where LLMs serve as judges (LaaJ). 
Prior studies have also leveraged LLMs to assist in identifying a generated-code bug's root causes~\cite{dou2024s}. 
In this context, we propose to leverage LLMs to assess the quality of generated code. 
Specifically, we employ the \textit{GPT-4o-mini} model, which offers performance comparable to GPT-4 while being more cost-effective~\cite{openai2024gpt4omini}. 
Regarding the hyperparameters, we considered a \textit{temperature} of {$\mathit{0}$ aiming to achieve a more conservative generation and \textit{max\_tokens} set to {$\mathit{5120}$ to ensure sufficient space for detailed output. 
 
Knowing that \textit{GPT-4} models can be shaped and optimized by setting system prompts, which define the model's persona~\cite{wang2024ai}, we adopted the persona of an \textit{expert in code quality analysis} with a focus on identifying specific inefficiencies in machine-generated code. 
Such a persona guides the model to prioritize practical and significant code inefficiencies rather than minor or irrelevant inefficiencies.  

Finally, we defined a fixed user prompt used as input when asking for judgment support (\textit{judger}). 
In this prompt, we input the generated code snippet and its task description, providing a context to aid the \textit{judger} in identifying logical inefficiencies in the generated code.
As a result of this step, we enriched the dataset $\mathit{B}$ with GPT-generated judgments, creating dataset $\mathit{C}$, as denoted in Figure \ref{fig:taxonomy_methodology}. 
Specifically, $\mathit{C}$ consists of 492 generated code samples, each paired with its corresponding LLM-generated judgment.

\subsection{Step 3 and 4: Manual Labeling}
\textit{Step 3: Manual Labeling - Pilot Study} 
With no predefined categories from previous studies, our objective is to construct a taxonomy of inefficiencies using an inductive, bottom-up approach through manual analysis of LLM-generated code snippets. 
To initiate this process, we randomly selected 196 samples $\mathit{C}$ (40\%, pilot set), ensuring an equal proportion of cases from each model.
The first two authors, a senior Master's student and a Postdoctoral Fellow with experience in Python and Software Engineering (SE) research, collaboratively analyzed and labeled the pilot set (40\% of samples) using open coding~\cite{seaman1999qualitative}, a method widely used for taxonomy construction in SE~\cite{yahmed2023deploying, 10.1145/3377811.3380395, 10.1145/3544548.3580817}.  
The analysis involved examining the task description, the generated code, code quality judgment, and the execution and testing status (from the unit tests available in HumanEval+). 
Without predefined assumptions, inefficiencies were iteratively identified and grouped into categories and subcategories.
Since LaaJ may hallucinate~\cite{sollenberger2024llm4vv}, we treated their judgments as a starting point and independently verified and expanded on the identified inefficiencies.  During this process, the labelers assigned brief, descriptive labels to each inefficiency observed in the pilot set. 
These initial labels were then categorized, forming a hierarchical taxonomy of inefficiency types. 
This iterative process involved continuously refining the categories as they moved back and forth between individual samples and emerging patterns, ensuring that the taxonomy accurately reflected the inefficiencies underlying the generated code. 
All tasks in the pilot set were thoroughly discussed, and the labelers reached a consensus on both categories and subcategories, resulting in a taxonomy comprising \textbf{$5$} inefficiency categories and \textbf{$18$} subcategories 
and a fully labeled pilot set. 
This step resulted in a codebook of inefficiency patterns, with multi-labeling allowed, since a single code snippet can exhibit multiple types of inefficiencies~\cite{tambon2024bugs}.

\textit{Step 4: Taxonomy Refinement \& Full Dataset Labeling}
After labeling the pilot set and constructing the initial codebook of inefficiencies, the labelers independently labeled the remaining 60\% samples, following the same procedure. 
If any generated code could not be categorized within the existing taxonomy, they were marked as \textit{Pending} and reviewed to determine whether new categories or subcategories were needed. Through these discussions, \textit{Syntax Error} was added as a subcategory under \textit{Errors}, leading to a refined taxonomy comprising five categories and 19 subcategories, providing a more comprehensive representation of inefficiencies in LLM-generated code.

Once all samples were labeled, we assessed inter-rater agreement to evaluate the consistency between the two labelers. Following similar studies~\cite{yahmed2023deploying, tambon2024bugs}, we used Cohen’s Kappa~\cite{doi:10.1177/001316446002000104}, calculated at category and subcategory levels. Inter-rater agreement (Cohen’s Kappa) reached 0.846 (for category) and 0.735 (for subcategory), aligning with prior SE studies~\cite{yahmed2023deploying, tambon2024bugs}.

\subsection{Step 5: Participants' feedback and survey analysis}

To evaluate the relevance and gather insights into the perceived popularity of the identified inefficiencies in real-world settings, we conducted a survey targeting software practitioners and researchers using LLMs in their code generation tasks. 
Following prior studies~\cite{fu2023security, tambon2024bugs}, we collected the email addresses of GitHub users who collaborated on repositories containing LLM-generated code. 
To identify relevant repositories, we searched GitHub for repositories where at least one file contained code generated by LLM-based tools or models, including Copilot, ChatGPT, Codex, CodeLlama, Llama, and Llama-2. Specifically, we used the following keyword-based queries: association between \texttt{by/with} and \texttt{Copilot, ChatGPT, Codex, CodeLlama, Llama, Llama-2}. This process yielded $411$ repositories.
Next, we extracted publicly available email addresses of contributors from these repositories using PyDriller~\cite{spadini2018pydriller}. This resulted in $835$ unique email addresses, of which $51$ were unreachable, leaving us with $784$ successfully delivered emails (mined on November 18, 2024). 

To ensure clear communication and maximize participant engagement, we structured the survey to clearly outline its objectives, structure, and expectations. 
First, the survey outlined a detailed message about its purpose, scope, and estimated completion time. 
The survey questionnaire is divided into two main parts. 
The first part consists of demographic and open-ended questions regarding the LLMs participants have used, as well as the inefficiencies they encountered based on their experiences.
The second part assessed the frequency and relevance of inefficiencies in LLM-generated code, as defined in Step 4. For each inefficiency category, we provided a general description, followed by its specific subcategories, each accompanied by a description and an example code snippet.
Participants were then asked two questions per subcategory, rated on a 5-point Likert scale: (i) how frequently they encountered the inefficiency reported and (ii) how important they believed it was to address it. 
Additionally, the survey included an optional question, allowing participants to report any inefficiency categories or subcategories not covered in the taxonomy, based on their experience. 
The survey remained open for $4$ weeks, and we sent a reminder email two weeks after the first call. Once the results for each inefficiency pattern were collected, we processed the results to evaluate their popularity and relevance. 
This was achieved by aggregating the results using weighted averages of Likert scale values, following methodologies from similar studies~\cite{tambon2024bugs}. 
Additionally, we analyzed participants' feedback to gain deeper insights into their experiences with LLM-generated code. Their comments provided context for interpreting the results and helped identify potential gaps or overlooked aspects in the taxonomy.

\section{Results}
\label{sec: results}
In this section, we present the findings for our two RQs explored in this study. We first introduce our taxonomy of inefficiencies in LLM-generated code, followed by an analysis of the perceptions of practitioners and researchers on the proposed taxonomy, as gathered through our online survey.

\subsection{\textbf{RQ1:} Patterns of Inefficiencies in LLM-generated Code}
\subsubsection{Taxonomy} We organized our observations as a taxonomy of five categories with 19 subcategories. 
The complete taxonomy is shown in Figure \ref{fig:taxonomy}. 
In the following, we explain each of the five categories with their respective subcategories. Examples of each inefficiency subcategory (i.e., code snippets) can be found in our replication package. 

\begin{figure*}[h!]
 \centering
  \includegraphics[width=\linewidth{}{}{}]{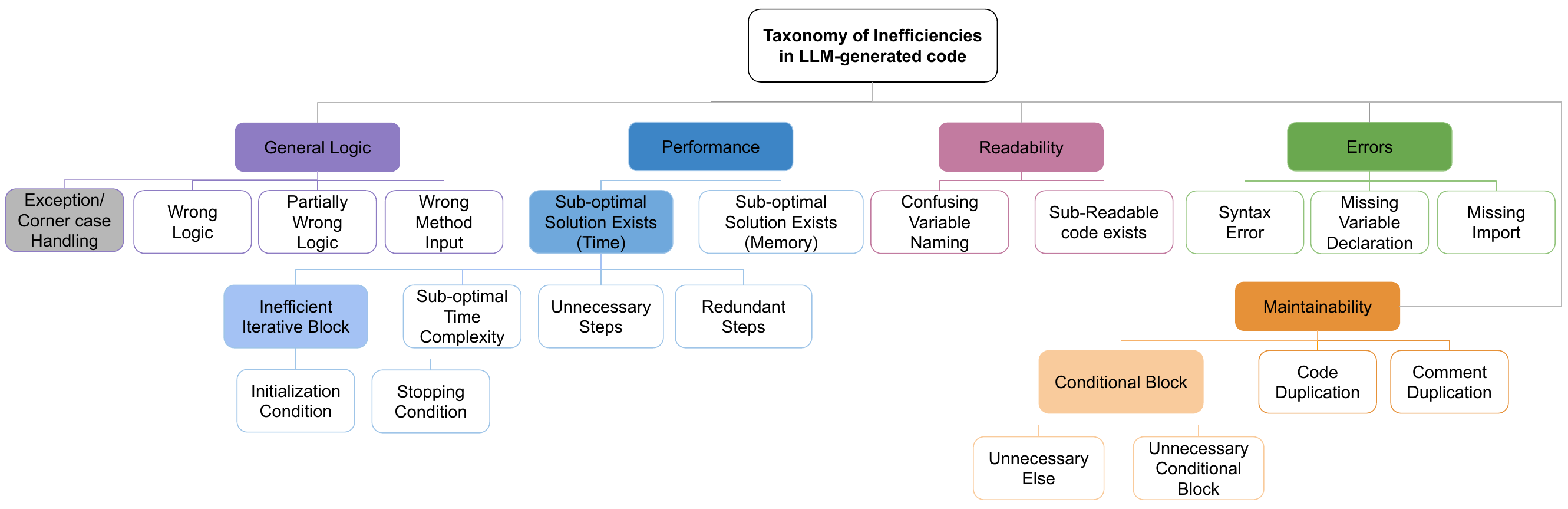}
  \caption{LLM-generated Code Inefficiencies: A comprehensive Taxonomy}
 \label{fig:taxonomy}
 \vspace{-1em}
\end{figure*}
\textbf{A. General Logic.} This category encompasses inefficiencies related to the general logic of a given task, including inefficiencies in understanding requirements and translating them into code.
\begin{itemize}
    \item \textbf{Wrong Logic.} It occurs when the generated code lacks meaningful logic, containing only comments, hard-coded values, returning method inputs, or using a simple \texttt{pass} statement. It also includes case-by-case implementations that do not follow a coherent logic. Additionally, it encompasses situations where all requirements are implemented incorrectly, with none being addressed properly.
    \item \textbf{Partially Wrong Logic} Different from \textit{Wrong Logic}, this subcategory occurs when some requirements requested in the prompt are implemented correctly while others are either implemented incorrectly or not implemented at all.
    \item \textbf{Wrong Method Input.} It occurs when the LLM generates a statement that is generally coherent with the logic but uses incorrect or inappropriate inputs. For example, passing a single element to the \texttt{sorted} method instead of a list may lead to an unintended behavior. 
    \item \textbf{Exception/Corner Case Handling.} This occurs when the generated code lacks exception handling or fails to address corner cases. 
    Initially, this subcategory was merged with \textit{Partially Wrong}. However, following the suggestions of the survey participants, it was detached as a standalone subcategory. 
\end{itemize}

\textbf{B. Performance.} This category encompasses inefficiencies that degrade the efficiency of the generated code, affecting aspects such as memory management and execution time. 

\begin{itemize}
    \item \textbf{Sub-Optimal Solution Exists (Memory)}
    This subcategory captures cases where the same logic can be implemented in a more memory-efficient way. For example, recursion may be avoided when intermediate results are stored on the call stack, potentially leading to memory overflow for deep recursions. Similarly, list comprehensions, which store all results in memory, can be replaced with generator expressions to enable lazy evaluation and reduce memory usage.
    \item \textbf{Sub-Optimal Solution Exists (Time)}
    It identifies the cases where there is an equivalent, more time-efficient algorithm compared to the LLM-generated one. In this subcategory, we report four additional subcategories: 
    \begin{itemize}
        \item \textit{Sub-Optimal Solution (Time Complexity)}: It occurs when the implemented logic can be replaced with another algorithm with lower time complexity. For instance, using \textit{Quick Sort}, with a time complexity of \(O(n \log n)\) on average, is generally more efficient than \textit{Insertion Sort}, which has a time complexity of \(O(n^2)\) in the average and worst cases —particularly when sorting large lists.
        \item \textit{Unnecessary Steps.} It refers to instructions in the code that do not affect or contribute to the final output. Eliminating such steps enhances the code's overall efficiency without altering its functionality, such as unnecessary type conversion.
        \item \textit{Redundant Steps.} This occurs when an essential instruction to the logic is repeated unnecessarily. To avoid such cases, the given instruction could be executed once, and its output could be stored in a local variable instead of re-executing the same instruction.
        \item \textit{Inefficient Iterative Block.} This category focuses on inefficiencies within repetitive structures, such as loops and iteration management.\\
            + \textit{Iterative Block (Initialization Condition).} It occurs when the initialization condition of a loop is too broad and can be narrowed without affecting the correctness of the method. For example, when determining whether a number is prime, we can stop checking the divisibility at \( \sqrt{n} \) rather than iterating all the way up to \( n \).\\
            + \textit{Iterative Block (Stopping Condition).} It occurs when the stopping condition for a repetitive block is very broad and can be optimized without changing the intended functionality or when it lacks an early stopping condition when applicable. For instance, when searching for an element in a sorted ascending list, instead of iterating through the entire list, we can stop early if the current element exceeds the target value. 
    \end{itemize}
\end{itemize}
\textbf{C. Readability} This category encompasses inefficiencies related to the overall readability, clarity, and ease of comprehensibility of LLM-generated code snippets.
\begin{itemize}
    \item \textbf{Confusing Variable Naming.} It refers to variable names that may confuse the developer, even if they do not result in runtime errors. It includes inefficiencies such as using existing identifiers to define new variables (e.g., using \texttt{dict} as a variable or method name) or variable shadowing, where the same variable name is used in different scopes. 
    \item \textbf{Sub-readable method exists.} This refers to cases where the code is difficult to understand or does not adhere to conventions. For example, the combination of list comprehension, \texttt{map}, and \texttt{lambda} functions, along with a for \texttt{loop} and multiple method calls, such as the use of this statement: 
    \texttt{return reduce(lambda x, y: x * y, [int(i) for i in str(n) if int(i) \% 2 == 1])}. 
\end{itemize}
\textbf{D. Maintainability.} This category highlights inefficiencies in LLM-generated code that may negatively impact the maintainability of the code, possibly intended to be integrated into larger systems.
\begin{itemize}
    \item \textbf{Code Duplication.} This occurs when identical or similar blocks of code appear in multiple places without contributing to the main logic, essentially resembling code cloning~\cite{ain2019systematic}. 
    \item \textbf{Comment Duplication.} It occurs when repetitive comments are used excessively, identical, or similar, without providing additional insight or explanations about the logic. 
    \item \textbf{Conditional Block.} It groups inefficiencies related to conditional blocks, i.e., \texttt{if} blocks. Although it does not affect the functionality, it is important to be addressed for better code clarity and easier testing/debugging~\cite{lacerda2020code}. While in our studied sample we did not observe \texttt{foreach} blocks, this subcategory may also be extended to cover them.
    \begin{itemize}
        \item \textit{Unnecessary Else.} It takes place when an \texttt{else} statement is used unnecessarily within an \texttt{if} block and can be removed without affecting the logic. It usually happens when the preceding \texttt{if} or \texttt{elif} statements already contain a \texttt{return} or \texttt{break}, making the \texttt{else} unnecessary.
        \item \textit{Unnecessary Conditional Block.} It refers to cases where using a conditional block adds unnecessary complexity, and the code can be simplified by directly using the output of a given condition in a \texttt{return} statement.
    \end{itemize}
\end{itemize}
\textbf{E. Errors} Inefficiencies falling under \textit{Erros} are related to the fundamental correctness of the code regardless of the logic being implemented, like runtime errors, unexpected behavior, or failure to execute the code properly. 
\begin{itemize}
    \item \textbf{Missing Import.} It occurs when methods or constants from external modules/libraries are used without being properly imported. Missing imports can lead to errors or failures when the code attempts to use the functionality provided by those modules. 
    \item \textbf{Missing Variable Declaration.} It occurs when a variable is used without being properly declared, initialized, or passed as an argument for a method, which may result in runtime errors. 
    \item \textbf{Syntax Error.} This occurs when the code contains a syntax error preventing it from being executed, such as improper use of keywords or incorrect indentations in Python. 
\end{itemize}

Overall, LLM-generated code exhibits similar inefficiencies to human-written code. 
While LLMs often pass more test cases than human developers, they struggle with tasks requiring domain expertise~\cite{licorish2025comparing}. 
Regarding performance, although LLMs can generate efficient solutions for certain tasks, they often under-perform compared to human-optimized implementations~\cite{coignion2024performance}. 
In terms of readability, LLM-generated code is generally comparable to human-written code but may deviate from coding standards~\cite{zheng2024beyond, takerngsaksiri2025code}. However, its increased complexity can elevate maintenance effort~\cite{licorish2025comparing}. 
While LLMs can address maintainability issues, their fixes frequently introduce new errors at a higher rate than human corrections~\cite{nunes2025evaluating}.

\subsubsection{Analysis of Inefficiency Patterns} 
\begin{table*}[]
\centering
\caption{(1) Frequency of inefficiencies in the studied sample of LLM-generated code, and (2) Survey-based taxonomy validation Results}
\begin{tabular}{c|cccc||cc|}
\cline{2-7}
                                                                 & \multicolumn{4}{c||}{\textit{\textbf{(1)Frequency in of ineffeciencies in LLM-generated code(\%)}}}                                                                                   & \multicolumn{2}{c|}{\textit{\textbf{\begin{tabular}[c]{@{}c@{}} (2) Survey-based Validation\\ (Weighted average)\end{tabular}}}} \\ \cline{2-7} 
                                                                 & \multicolumn{1}{c|}{\textit{Codellama}} & \multicolumn{1}{c|}{\textit{Deeepseek-Coder}} & \multicolumn{1}{c|}{\textit{CodeGemma}} & \textit{Overall} & \multicolumn{1}{c|}{\textit{Popularity}}                                & \textit{Relevance}                                \\ \hline
\multicolumn{1}{|c|}{\textit{\textbf{1- General Logic}}}         & \multicolumn{1}{c|}{60.37}              & \multicolumn{1}{c|}{70.73}                    & \multicolumn{1}{c|}{\textbf{74.39}}              & 68.5             & \multicolumn{1}{c|}{---}                                                & 4.03                                              \\ \hline
\multicolumn{1}{|c|}{Wrong Logic}                                & \multicolumn{1}{c|}{9.76}               & \multicolumn{1}{c|}{39.63}                    & \multicolumn{1}{c|}{\textbf{57.93}}              & \textbf{35.77}            & \multicolumn{1}{c|}{2.22}                                               & 3.64                                              \\ \hline
\multicolumn{1}{|c|}{Partially Wrong Logic}                      & \multicolumn{1}{c|}{\textbf{47.56}}              & \multicolumn{1}{c|}{27.44}                    & \multicolumn{1}{c|}{15.24}              & 30.08            & \multicolumn{1}{c|}{3.17}                                               & 3.88                                              \\ \hline
\multicolumn{1}{|c|}{Wrong Method Input}                         & \multicolumn{1}{c|}{\textbf{4.27}}               & \multicolumn{1}{c|}{3.66}                     & \multicolumn{1}{c|}{1.22}               & 3.05             & \multicolumn{1}{c|}{2.76}                                               & 3.53                                              \\ \hline
\multicolumn{1}{|c|}{\textit{\textbf{2- Performance}}}           & \multicolumn{1}{c|}{\textbf{48.78}}              & \multicolumn{1}{c|}{30.49}                    & \multicolumn{1}{c|}{23.17}              & 34.15            & \multicolumn{1}{c|}{---}                                                & 3.52                                              \\ \hline
\multicolumn{1}{|c|}{Sub-Optimal Solution (Memory)}              & \multicolumn{1}{c|}{\textbf{26.83}}              & \multicolumn{1}{c|}{17.07}                    & \multicolumn{1}{c|}{12.2}               & \textbf{18.70}             & \multicolumn{1}{c|}{3.07}                                               & 3.12                                              \\ \hline
\multicolumn{1}{|c|}{Iterative Block (Initialization Condition)} & \multicolumn{1}{c|}{0.61}               & \multicolumn{1}{c|}{0.61}                     & \multicolumn{1}{c|}{0.61}               & 0.61             & \multicolumn{1}{c|}{\multirow{2}{*}{2.69}}                              & \multirow{2}{*}{3.19}                             \\ \cline{1-5}
\multicolumn{1}{|c|}{Iterative Block (Stopping Condition)}       & \multicolumn{1}{c|}{\textbf{3.66}}               & \multicolumn{1}{c|}{1.22}                     & \multicolumn{1}{c|}{2.44}               & 2.44             & \multicolumn{1}{c|}{}                                                   &                                                   \\ \hline
\multicolumn{1}{|c|}{Redundant Steps}                            & \multicolumn{1}{c|}{\textbf{9.76}}               & \multicolumn{1}{c|}{3.05}                     & \multicolumn{1}{c|}{3.05}               & 5.28             & \multicolumn{1}{c|}{2.64}                                               & 3.05                                              \\ \hline
\multicolumn{1}{|c|}{Unnecessary Steps}                          & \multicolumn{1}{c|}{\textbf{4.27}}               & \multicolumn{1}{c|}{2.44}                     & \multicolumn{1}{c|}{0.61}               & 2.44             & \multicolumn{1}{c|}{3.03}                                               & 3.03                                              \\ \hline
\multicolumn{1}{|c|}{Sub-Optimal Time Complexity}                & \multicolumn{1}{c|}{\textbf{23.78}}              & \multicolumn{1}{c|}{17.68}                    & \multicolumn{1}{c|}{14.02}              & 18.5             & \multicolumn{1}{c|}{2.93}                                               & 3.12                                              \\ \hline
\multicolumn{1}{|c|}{\textit{\textbf{3-Readability}}}            & \multicolumn{1}{c|}{\textbf{6.71}}               & \multicolumn{1}{c|}{6.10}                      & \multicolumn{1}{c|}{1.22}               & 4.67             & \multicolumn{1}{c|}{---}                                                & 3.22                                              \\ \hline
\multicolumn{1}{|c|}{Confusing Variable Naming}                  & \multicolumn{1}{c|}{1.22}               & \multicolumn{1}{c|}{1.22}                     & \multicolumn{1}{c|}{0.61}               & 1.02             & \multicolumn{1}{c|}{2.17}                                               & 3.00                                                 \\ \hline
\multicolumn{1}{|c|}{Sub-Readable Code Exists}                   & \multicolumn{1}{c|}{\textbf{5.49}}               & \multicolumn{1}{c|}{4.88}                     & \multicolumn{1}{c|}{0.61}               & \textbf{3.66}             & \multicolumn{1}{c|}{2.79}                                               & 3.05                                              \\ \hline
\multicolumn{1}{|c|}{\textit{\textbf{4-Maintainability}}}        & \multicolumn{1}{c|}{\textbf{29.88}}              & \multicolumn{1}{c|}{12.8}                     & \multicolumn{1}{c|}{20.73}              & 21.14            & \multicolumn{1}{c|}{---}                                                & 3.31                                              \\ \hline
\multicolumn{1}{|c|}{Unnecessary Else}                           & \multicolumn{1}{c|}{\textbf{21.95}}              & \multicolumn{1}{c|}{10.98}                    & \multicolumn{1}{c|}{19.51}              & \textbf{17.48}            & \multicolumn{1}{c|}{2.29}                                               & 2.59                                              \\ \hline
\multicolumn{1}{|c|}{Unnecessary Conditional Block}              & \multicolumn{1}{c|}{3.66}               & \multicolumn{1}{c|}{1.83}                     & \multicolumn{1}{c|}{3.66}               & 3.05             & \multicolumn{1}{c|}{2.62}                                               & 2.78                                              \\ \hline
\multicolumn{1}{|c|}{Code Duplication}                           & \multicolumn{1}{c|}{\textbf{4.88}}               & \multicolumn{1}{c|}{1.83}                     & \multicolumn{1}{c|}{0.61}               & 2.44             & \multicolumn{1}{c|}{2.46}                                               & 3.02                                              \\ \hline
\multicolumn{1}{|c|}{Comment Duplication}                        & \multicolumn{1}{c|}{\textbf{3.05}}               & \multicolumn{1}{c|}{0.00}                        & \multicolumn{1}{c|}{0.61}               & 1.22             & \multicolumn{1}{c|}{1.86}                                               & 2.10                                              \\ \hline
\multicolumn{1}{|c|}{\textit{\textbf{5-Errors}}}                 & \multicolumn{1}{c|}{\textbf{7.32}}               & \multicolumn{1}{c|}{6.1}                      & \multicolumn{1}{c|}{3.66}               & 5.69             & \multicolumn{1}{c|}{---}                                                & 3.93                                              \\ \hline
\multicolumn{1}{|c|}{Syntax Error}                               & \multicolumn{1}{c|}{0.00}                  & \multicolumn{1}{c|}{2.44}                     & \multicolumn{1}{c|}{2.44}               & 1.63             & \multicolumn{1}{c|}{2.29}                                               & 3.79                                              \\ \hline
\multicolumn{1}{|c|}{Missing Variable Declaration}               & \multicolumn{1}{c|}{0.61}               & \multicolumn{1}{c|}{0.00}                        & \multicolumn{1}{c|}{0.61}               & 0.41             & \multicolumn{1}{c|}{2.17}                                               & 3.60                                              \\ \hline
\multicolumn{1}{|c|}{Missing Module Import}                      & \multicolumn{1}{c|}{\textbf{6.10}}                & \multicolumn{1}{c|}{3.66}                     & \multicolumn{1}{c|}{1.22}               & \textbf{3.66}             & \multicolumn{1}{c|}{2.86}                                               & 3.26                                              \\ \hline
\end{tabular}
\label{table:benchmark_frequency_survey_popularity_relevance}
\vspace{-1em}
\end{table*}
Most samples were labeled with a single category (56.5\%), followed by those with two (24.59\%) and 9.96\% were not labeled, as raters found no inefficiencies and considered them efficient.
A similar distribution appeared at the subcategory level with 54.07\% of samples assigned one subcategory followed by 18.29\% with two and 11.38\% with three.  
Notably, 33.54\% of the samples spanned multiple categories, highlighting the interconnected nature of inefficiency patterns across different code quality aspects.

Table \ref{table:benchmark_frequency_survey_popularity_relevance} presents the frequency of inefficiency categories and subcategories (since multi-labeling is allowed, totals may exceed 100\%). 
\textit{General Logic} emerges as the most frequent category across all models (68.5\% of samples). 
Among the selected models, CodeGemma, the least efficient, based on Top@k (leaderboard~\cite{evalplusLeaderboard}), exhibited the highest rate of logic-related inefficiencies (74.39\%), often generating code without a meaningful logic (57.93\%). 
While the most efficient model based on Top@K, CodeLlama, showed fewer critical logic errors (60.37\%), though it also frequently generated partially correct solutions (47.56\%), suggesting that even advanced models may misinterpret requirements, which is aligned with findings of other research works~\cite{xu2025large}.

\textit{Performance} inefficiencies ranked second (34.15\%), with CodeLlama exhibiting the highest frequency within this category.
\textit{Sub-Optimal (Memory)} (18.7\%), \textit{Sub-Optimal (Time Complexity)} (18.5\%) and \textit{Redundant Steps} (5.28\%)  were the most prevalent performance-related inefficiencies.
These inefficiencies are not exclusive to smaller or open-source models; prior studies have reported similar performance issues in GPT-4-generated code~\cite{huang2024effibench}, reinforcing the broader challenge of optimizing LLM-generated code across different model architectures.

\textit{Maintainability} inefficiencies varied significantly between models. DeepSeek-Coder, trained exclusively on code, exhibited the lowest \textit{Maintainability} issues (12.8\%), whereas CodeLlama and CodeGemma (trained on mixed code/natural language data) have higher rates (29.88\% and 20.73\%, respectively). 
This discrepancy suggests that code-specific training may improve structural quality, while incorporating natural language data in training may dilute the focus on maintainability. 
The prevalence of \textit{Unnecessary Else} blocks (21.95\% in CodeLlama) underscores a broader challenge: LLMs often generate unnecessarily complex control flows, which complicate long-term maintenance~\cite{fischer2023designing}. 

\textit{Readability} (4.67\%) and \textit{Errors} (5.69\%) were less frequent but non-trivial. \textit{Syntax Errors} were rare ($<2.44\%$), confirming the strong syntactic grasp of code LLMs. However, Missing Imports (6.1\% in CodeLlama) reveal persistent gaps in dependency resolution, a known limitation even in highly-efficient models like GPT-4~\cite{wang2024and}.

\begin{figure}[t]
  \centering
  \includegraphics[width=1\columnwidth]{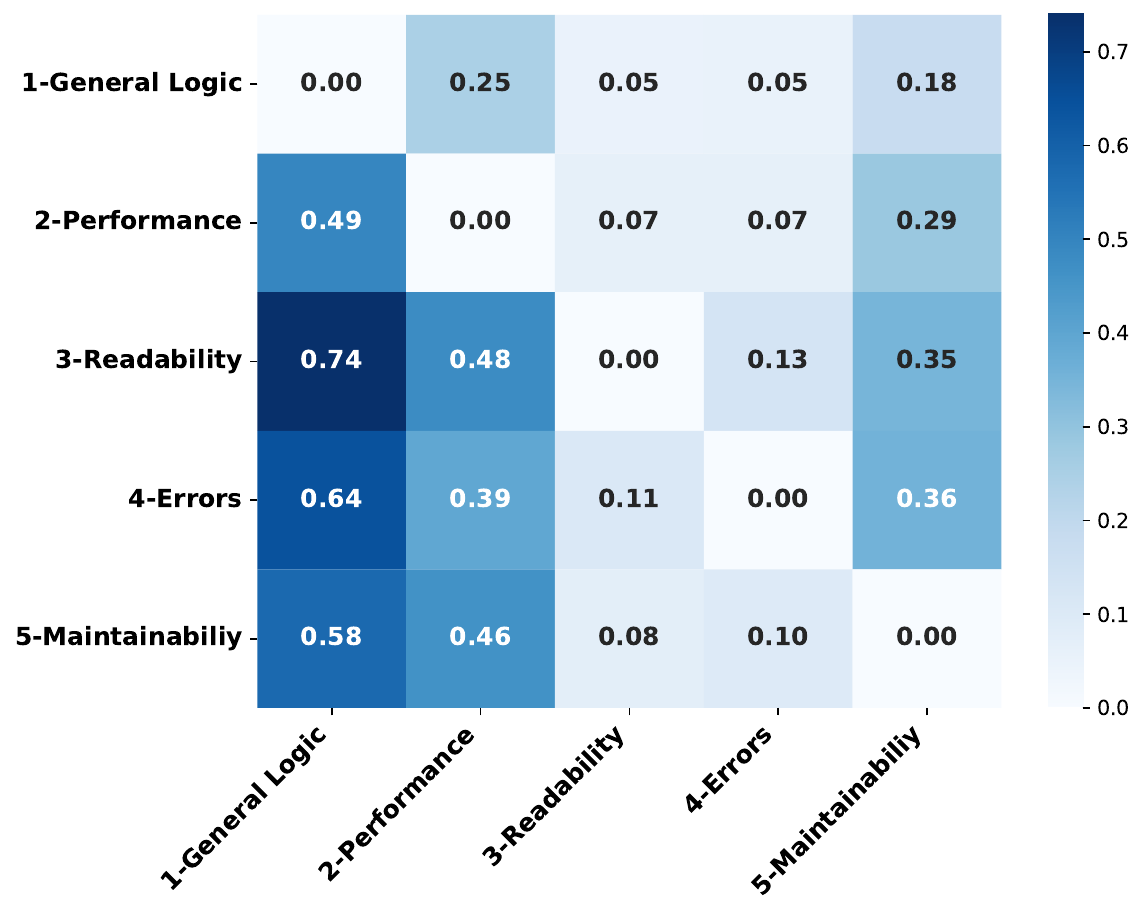}
  \caption{Normalized Category Co-occurrence} 
  \label{fig:all_categories_heat_map}
  \vspace{-1em}
\end{figure}
To gain deeper insights into inefficiency relationships, we analyze how different categories co-occur across models.
Figure \ref{fig:all_categories_heat_map} presents the normalized category 
co-occurrence heatmap across all models. 
For each inefficiency category, we calculated the number of times it appeared together with every other category within the same code snippet. These counts were then summed across all models and normalized row-wise by the total occurrences of the corresponding category.

Our results reveal strong inter-dependencies between inefficiency patterns in LLM-generated code. \textit{General Logic} inefficiencies exhibit the highest co-occurrence rates with other categories, particularly \textit{Readability (0.74)}, \textit{Errors (0.64)}, and \textit{Maintainability (0.58)}. This suggests that flawed logic not only impacts correctness but also contributes to inefficiencies in clarity, structure, and long-term maintainability of the generated code. Similarly, \textit{Performance} inefficiencies frequently co-occur with \textit{Readability (0.48)} and \textit{Maintainability (0.46)}, highlighting that performance inefficiencies are often tied to code complexity and structure, rather than isolated algorithmic inefficiencies.
Interestingly, \textit{Errors} demonstrate weaker co-occurrences with other inefficiencies, with their strongest link being to \textit{Maintainability (0.36)}, indicating that while syntax and dependency errors are less common, they may still hinder maintainability efforts by requiring additional debugging and refactoring. Meanwhile, \textit{Readability} and \textit{Maintainability} are closely related (0.35), reinforcing the idea that clear code structures contribute to long-term maintainability.
These findings suggest that improving LLM-generated logic and performance could lead to broader improvements across multiple inefficiency categories, reducing the need for manual corrections and enhancing code usability in real-world development.

\subsection{RQ2: Relevance of Identified Inefficiencies for Practitioners and Researchers}
The survey remained open for four weeks, reaching $58$ responses, resulting in a response rate of $7.39$\%, which is in line with rates observed in similar surveys in Software Engineering 
\cite{yahmed2023deploying, nardone2023video, nikanjam2022faults, serban2022adapting}. 
First, we present the demographic information about the participants and their experience with LLM-assisted coding. 
Then, we provide a detailed analysis of the popularity and perceived relevance of inefficiencies in LLM-generated code reported by the participants.

\subsubsection{Demographic Information and Experience with Coding using LLMs}
All participants answered the demographic questions. We had participants from both academia (60\%) and industry (40\%).
The academic participants consisted of graduate students (29), undergraduate students (three), and postdoctoral researchers/lecturers (three). 
Industry participants included 11 software developers, seven ML/data science developers, four senior executives, and one cybersecurity developer. 
About coding experience, 37 participants had more than 5 years of experience, while 11 had between 3-5 years, and 10 had between 1-3 years.
LLM-assisted coding is widely adopted, with 98\% of the participants using LLMs for coding. 
Among them, 68\% rely exclusively on proprietary models, with ChatGPT and Copilot being the most popular (89.66\% and 29.31\%, respectively). 
Regarding open-source, the most popular models are Mistral, the Phi model family, and Qwen-Coder (5.17\%, 3.45\%, and 3.45\%, respectively).
Participants reported using a variety of programming languages, with the top five being Python, C/C++, JavaScript, Java, and Go (91.38\%, 18.97\%, 18.97\%, 8.62\%, and 3.45\%, respectively). 

Before the questions about our taxonomy, we asked participants about their general experiences and challenges with LLM-generated code. 
Reported challenges can be categorized into two main groups: \textit{user experience} (outside the scope of our taxonomy) and \textit{code-related}, which are directly aligned with our research.
We discuss the inefficiencies encountered by the participants based on their experiences before seeing the taxonomy.

\paragraph{User Experience-Related Inefficiencies} 
These inefficiencies focus on the challenges/difficulties that users face when interacting with LLMs, rather than the generated code itself. \textit{Six} participants reported that LLMs struggle with debugging or modifying specific parts of code, a finding consistent with previous studies~\cite{tian2024debugbench}.
In the same way, programming language and library popularity also play a significant role in the efficiency of LLM-generated code. 
\textit{five} participants noted that LLMs generate more efficient code for popular languages and libraries like Python and NumPy compared to less common ones like DHL and Tinkter. This difference is likely because LLMs are trained on more examples of popular languages and libraries, making them more familiar with their syntax and usage. 
Programming language paradigms also play a role—P12\footnote{We refer to survey participants using the notation Px, where x is an assigned participant ID.} observed that LLMs handle interpreted languages better than compiled ones.
Finally, participants expressed varying preferences for chat-based LLMs (e.g., ChatGPT) versus inline code assistants (e.g., Copilot).
While P41 found Copilot suggestions as ``stupid suggestions and distracted'', P11 preferred inline assistance over chat-based interactions. 

\paragraph{Code-Related Inefficiencies} 
\label{p:code_inefficiencies}  
Participants also reported various inefficiencies in the generated code that align with the focus of our taxonomy. 
\textbf{19} participants stated that LLMs often struggle to produce the correct logic on the first attempt, even for simple tasks with detailed descriptions, requiring multiple further iterations. P22 specifically noted that LLMs may generate ``unnecessarily long or complex code (for simple tasks)''
However, P15 found that for more complex tasks such as building a Flask application, LLMs tend to perform ``oversimplification to the point of uselessness'', where generation is limited to high-level steps or generalized code that lack specificity, performance efficiency, and functionality. 
Additionally, \textit{three} participants reported that handling exceptions and corner cases is often overlooked. 
These reported inefficiencies fall under the \textit{General Logic} category of our taxonomy. 

Another notable challenge reported by the participants is the misalignment between LLMs' advancements and technological progress. \textbf{13} participants reported that LLMs struggle with version tracking, frequently generating code that uses deprecated methods, mixes library versions, or is incompatible with specified APIs (also mentioned by previous works~\cite{wang2024and}).
These inefficiencies align with \textit{Wrong Method Input} (\textit{General Logic}) and \textit{Missing Import} (\textit{Errors}) subcategories of our taxonomy.

\textit{Readability} inefficiencies were also reported by 7 respondents spanning various programming languages, like Python, C/C++, Java, JavaScript, Go, and Bash. They highlighted poor formatting, failure to adhere to coding conventions, inconsistent variable names, and reliance on outdated practices as reported by Zhang et al.~\cite{zheng2024beyond}.

Additionally, \textit{Six} participants highlighted performance issues. Others pointed to a lack of robust, modular, object-oriented architecture and insufficient documentation and comments as well. 
The inefficiencies reported by the participants align with the literature~\cite{zheng2024beyond, dou2024s, zheng2024beyond} and fall under \textit{Readability}, \textit{Performance}, and \textit{Maintainability} categories of our taxonomy.

Besides inefficiencies, \textit{Seven} participants identified context limitations as a key issue. \textit{Two} participants found that LLMs had a limited context window and frequently forgot past interactions, as mentioned in related work~\cite{coignion2024performance}. 
Meanwhile, P5 reported that LLMs tended to retain outdated context, stating that they needed to ``start a new chat to get rid of hallucinations.''

\subsubsection{Popularity and Relevance of Code Inefficiencies: Survey Perception}
Table \ref{table:benchmark_frequency_survey_popularity_relevance} (6th and 7th columns) summarizes the aggregated results of popularity and relevance of the subcategories of code inefficiency patterns on a scale of 1 to 5. 
The \textit{popularity} score reflects how frequently participants faced each type of inefficiency, while \textit{relevance} aims to indicate whether the participants consider that such an inefficiency should be prioritized.
Results show that the frequency of inefficiencies in the studied sample set is aligned with popularity scores reported by participants. 

\textit{General Logic} inefficiencies were rated as the most relevant category (4.03/5). In particular, \textit{Partially Wrong Logic} stands out as the most relevant and popular subcategory encountered by participants (3.88/5 and 3.17/5, respectively) and is highly frequent in our studied sample set (30.08\%). 
This observation suggests that while LLMs generate code that often aligns with task descriptions, they frequently miss certain requirements, requiring manual effort to refine the logic. 
Similar observations were also reported in other studies, particularly those analyzing Copilot-generated code~\cite{MORADIDAKHEL2023111734}. 
Although \textit{Wrong Logic} was reported as the most frequent subcategory in our studied sample (35.77\%), it receives a relatively low popularity score (2.22/5), which can be explained by the fact that participants mostly use proprietary models, with 98\% relying on at least one such model.
Since proprietary models generally perform better than open-sourced ones, participants may encounter \textit{Wrong Logic} less frequently in their own experience. 


\textit{Performance} inefficiencies, with a frequency of 34.15\% in our studied sample, were also considered highly relevant (3.52/5) and widely encountered by participants with a popularity score (~3.0/5). 
The most concerning inefficiencies were \textit{Sub-Optimal (Memory)} and \textit{Sub-Optimal (Time Complexity)} (~3.0/5 for popularity, 3.12/5 for relevance), as well as \textit{Inefficient Iterative Blocks}, which directly impact execution time. 
This finding suggests that while correctness is prioritized, performance inefficiencies in LLM-generated code can still add to developers' workload, as highlighted by other researchers~\cite{MORADIDAKHEL2023111734}.

For \textit{Maintainability}, participants reported a moderate popularity (2.3/5) but a high relevance score (3.31/5), indicating that while maintainability inefficiencies are encountered less frequently, respondents still considered them important.
\textit{Unnecessary Else}, the most frequently observed maintainability inefficiency in our studied sample (17.48\%), had the lowest relevance score (2.59/5), as many developers considered it a minor issue that does not affect correctness. In contrast, \textit{Code Duplication}, despite being the least frequent inefficiency (2.44\%) and less popular (2.46/5), was rated more relevant (3.02/5) due to its impact on maintenance effort and complexity of refactoring. This suggests that developers prioritize maintainability inefficiencies that directly increase long-term maintenance costs, even if they occur less often.

\textit{Readability} inefficiencies were relatively rare (4.67\% frequency) but still perceived as moderately relevant (3.22/5.0) despite low popularity scores. This observation highlights that even infrequent readability inefficiencies are critical to be addressed, as developers typically resolve them manually rather than relying fully on automated fixes~\cite{oliveira2021recommending}.
Finally, while \textit{Errors} were the least frequent (5.69\%) and the least popular (~2.25/5) inefficiencies encountered by participants, they had a high relevance score (3.93/5.0). The participants emphasized that, although rare, these errors must be addressed, as they prevent the code from being compiled or executed. 


Our findings suggest that the inefficiencies identified in our analysis accurately reflect the challenges practitioners and researchers encounter when dealing with LLM-generated code.

\subsubsection{LLM-generated Code Inefficiencies Taxonomy: Feedback and Completeness}
We gathered participants' feedback through open questions for this part.
Regarding \textit{General Logic}, participants noted that \textit{Wrong Logic} often occurs when tasks are complex or involve new/unfamiliar technologies. 
For \textit{Partially Wrong Logic}, participants acknowledged that they often correct these inefficiencies but emphasized that LLMs sometimes \textit{``generate buggy solutions, and debugging takes longer than writing the code from scratch''}(P41)
Additionally, some participants pointed out that partially correct logic can be difficult to detect, increasing the risk of unnoticed errors at runtime. P2 was concerned that: \textit{``A junior coder who uses this could cause issues if pushed to production code without proper review''}. 
For \textit{Performance}, participants from both academia and industry noted that they do not rely on LLMs for generating highly optimized code. Instead, they prioritize correctness, ensuring the code is bug-free, and manually optimize performance only when necessary.
Regarding \textit{Readability}, participants were indifferent to the clarity and structure as long as the code functioned correctly and could be manually refined if needed. 
For \textit{Maintainability}, participants noted that inefficiencies like ``code and comment duplication'' are less noticeable when using tools such as Copilot~\cite{microsoft_copilot} and Cursor AI~\cite{cursor_ai}. 
\textit{Unnecessary Else} statements were considered low-impact, and participants showed little interest in addressing them. 
For \textit{Errors}, while they can lead to runtime failures, participants found them easy to detect using IDEs and correct. 

Participants acknowledged that prompting can help mitigate inefficiencies related to \textit{General Logic}, \textit{Performance}, and \textit{Readability}. However, certain inefficiencies, such as \textit{Wrong Method Input}, still require developer expertise to resolve effectively.
Overall, participants agreed that the proposed taxonomy effectively captures inefficiencies in LLM-generated code, evidenced by its high popularity and relevance scores. Many of these inefficiencies were already mentioned by participants before they were introduced to our taxonomy (\ref{p:code_inefficiencies}), demonstrating a strong coverage. However, three participants suggested introducing a new subcategory for \textit{Exception and Corner Case Handling}, as these inefficiencies were not explicitly included in our taxonomy. 




\section{Related Work}
\label{sec: related_work}
Prior research has explored various aspects of LLM-generated code, identifying multiple inefficiencies, including those related to performance and readability. 
However, most studies categorizing inefficiencies focus on analyzing buggy code, limiting their scope to correctness issues. 
To bridge this gap, we introduce a systematic categorization of inefficiencies that extends beyond correctness, addressing boarder code aspects.

Recent studies have systematically assessed inefficiencies in LLM-generated code—spanning readability, maintainability, and performance—by leveraging automated tools such as static analysis, code execution frameworks, and quantitative metrics like pass@k, code complexity, and lines of code.
In terms of readability, LLM-generated code often deviates from standard coding best practices and conventions, leading to reduced clarity and long-term maintainability inefficiencies ~\cite{zheng2024beyond, licorish2025comparing}. 
Although LLMs can generate shorter solutions for complex tasks, their generated code often exhibits higher structure complexity, making them more difficult to read, maintain, and debug \cite{dou2024s}. 
Additionally, LLM-generated code are susceptible to various code smells, which may harden further maintenance ~\cite{ouedraogo2024test}.
Maintainability issues are also reflected by the occurrence of \textit{improper documentation, unused or undefined variables, and security vulnerabilities}~\cite{siddiq2024quality}.
LLM-generated code further raises safety concerns as they are prone to critical vulnerabilities, including several from the Common Weakness Enumeration (CWE) Top 25, reinforcing concerns that LLM-generated code may introduce significant security flaws if not carefully reviewed~\cite{fu2023security}.
Performance inefficiencies present an additional challenge. Although LLMs can generate functionally correct solutions, the generated code often lacks efficiency, leading to suboptimal execution times and increased resource consumption~\cite{MORADIDAKHEL2023111734, coignion2024performance}.

Other works opted for manual analysis for deeper investigation to identify existing inefficiencies. 
Tambon et al.~\cite{tambon2024bugs} analyzed buggy open-source LLM-generated code and introduced a taxonomy of bug patterns, while Dou et al.~\cite{dou2024s} categorized the root causes of common bugs.
We observe that our proposed taxonomy intersects with prior taxonomies primarily in two categories, along with their respective subcategories: \textit{General Logic} and \textit{Errors}, both of which contribute to buggy code. 
However, inefficiencies related to \textit{Performance}, \textit{Readability}, and \textit{Maintainability} have received less attention, as they do not always result in bugs but can still degrade overall software quality.
By addressing these overlooked inefficiencies, our taxonomy provides a more comprehensive inefficiency categorization, offering new insights into the challenges of integrating LLM-generated code into real-world development.

\section{Threats to Validity}
\label{sec: threats_validity}
\textit{Construct Validity.} The process of collecting and labeling generated code may influence our study's results and findings. 
While our methodology aligns with existing works proposing taxonomies in SE~\cite{tambon2024bugs, yahmed2023deploying}, we mitigate such a risk by clearly explaining our approach in this paper to ensure external validation. 
The use of LaaJ introduces a potential threat, as LLMs may produce incorrect or inconsistent judgments. To address this risk, we treated the LLM's judgments as an initial reference rather than a definitive verdict. The two labelers independently reviewed and verified all judgments to minimize inaccuracies and mitigate hallucinations.
In the absence of LLM-generated code inefficiency categorization, we recognized the risk of introducing bias into our classification. To address this, we adopted an open coding procedure and utilized different LLM models to enhance the robustness and reliability of our findings. Additionally, we surveyed practitioners and researchers to validate the completeness of our taxonomy. 

\textit{Internal Validity.} A key internal threat to validity is the potential bias in the manual labeling of LLM-generated code. 
To mitigate such a threat, the first two authors independently labeled the generated code snippets. 
The inter-rater agreement at the category level is  0.846, which aligns with similar studies in software engineering~\cite{yahmed2023deploying, tambon2024bugs}.
Additionally, to further validate our findings, we surveyed 58 participants from both academia and industry, ensuring a broader perspective on the identified inefficiencies. 
Additionally, we leveraged an LaaJ to assess generated code snippets, providing an additional layer of validation and reducing potential human bias in the labeling process.
However, we recognize that LaaJ judgments may have missed certain inefficiencies or introduced biases that could influence human labeling. To mitigate this risk, the two labelers independently reviewed, refined, and expanded upon the LLM's judgments, ensuring that inefficiencies were identified beyond those initially flagged by the model.
Finally, we used GPT-4o\texttt{-mini} to provide initial judgments on code inneficiencies. While all outputs were manually validated to ensure accuracy, the use of such an LLM for initial assessment could have introduced biases or inconsistencies. We mitigated such a threat by reviewing and correcting all code inefficiencies before including them in the final results.

\textit{External Validity.} One external threat to validity concerns the selection of LLMs used to generate code. 
In this study, we utilize three open-source LLMs: CodeLlama, DeepSeek-Coder, and CodeGemma, all of which have demonstrated efficiency and have been employed in prior studies to leverage the capabilities of LLMs for code generation~\cite{Szalontai2024CodeLlama, lei2024autocoder, poudel2024documint}
Another potential threat stems from the complexity of the benchmark tasks, as they may not fully represent real-world programming scenarios, which could limit the generalizability of the inefficiencies identified. 
To mitigate this threat, we use the HumanEval+ dataset, a widely adopted benchmark for evaluating LLM-generated code~\cite{ouyang2023llm, lin2024llm, miah2024user}. 
Such a benchmark consists of human-crafted algorithmic tasks, which, while not encompassing the full breadth of software development tasks, represent a fundamental component of real-world programming. 
Furthermore, our study focuses exclusively on Python code, the most widely used programming language. 
However, this scope introduces a limitation, as we may have overlooked inefficiencies that are specific to other programming languages. 
Future work could extend our analysis to additional programming languages to enhance generalizability. 
Another critical threat is the long-term relevance of our proposed taxonomy, given the rapid evolution of LLMs. 
As these models continue to improve, some inefficiencies may become less prevalent while new ones may emerge. 
To address this threat, future research should periodically reassess and refine the taxonomy to ensure its continuous applicability in evaluating LLM-generated code.

\textit{Conclusion Validity.} 
Some types of inefficiencies may have been overlooked, potentially affecting the conclusions of this paper. To mitigate this threat, we manually inspected 492 code samples generated by three different LLM models. To minimize errors, two annotators independently labeled each sample, followed by a discussion to resolve discrepancies. Finally, the results were validated through a survey.
We also provide a replication package to support the reproducibility of our findings \cite{ReplicationPackage2025}. 


\section{Conclusion}
\label{sec: conclusion}
In this study, we systematically investigated inefficiencies in LLM-generated code and introduced a taxonomy categorizing them into 5 categories: \textit{General Logic}, \textit{Performance}, \textit{Readability}, \textit{Maintainability}, and \textit{Errors}, with 19 specific subcategories. 
We validated the taxonomy through a survey of 58 practitioners and researchers, who largely confirmed its completeness and affirmed the relevance and prevalence of the identified inefficiencies in real-world scenarios.
Our findings reveal a critical gap in the ability of LLMs to generate efficient code. 
\textit{General Logic} and \textit{Performance} inefficiencies emerged as the most frequent and relevant inefficiencies, often co-occurring with \textit{Maintainability} and \textit{Readability} highlighting their impact on overall LLM-generated code quality.
The proposed taxonomy provides a structured basis for evaluating LLM-generated code, facilitating efficiency comparisons, and guiding future improvements in LLMs for code generation. 

\section{Acknowledgment}
This work is partially supported by the Fonds de Recherche du Quebec (FRQ), the Canadian Institute for Advanced
Research (CIFAR), the Natural Sciences and Engineering Research Council of Canada (NSERC), and the Canada Research Chairs Program. The authors would like to thank all participants who contributed to the survey for their valuable insights. The views and conclusions expressed in this paper are solely those of the authors and do not necessarily reflect those of the supporting organizations.

\balance
\bibliographystyle{IEEEtran}
\bibliography{refs}

\end{document}